# Standard Model CP-violation and Baryon asymmetry

M.B. Gavela,[1] P. Hernández[2], J. Orloff[3], O. Pène[4]

[1] CERN, TH Division, CH-1211, Geneva 23, Switzerland
[2] Lyman lab., Harvard University, Cambridge, MA 02138[1]
[3] Institut für Theoretische Physik, Univ. Heidelberg
[4] LPTHE, F 91405 Orsay, France,[2]

### Abstract

Simply based on CP arguments, we argue against a Standard Model explanation of the baryon asymmetry of the universe in the presence of a first order phase transition. A CP-asymmetry is found in the reflection coefficients of quarks hitting the phase boundary created during the electroweak transition. The problem is analyzed both in an academic zero temperature case and in the realistic finite temperature one. The building blocks are similar in both cases: Kobayashi-Maskawa CP-violation, CP-even phases in the reflection coefficients of quarks, and physical transitions due to fermion self-energies. In both cases an effect is present at order $\alpha_W^2$ in rate. A standard GIM behaviour is found as intuitively expected. In the finite temperature case, a crucial role is played by the damping rate of quasi-particles in a hot plasma, which is a relevant scale together with $M_W$ and the temperature. The effect is many orders of magnitude below what observation requires, and indicates that non standard physics is indeed needed in the cosmological scenario.

---

[1] Junior Fellow, Harvard Society of Fellows
[2] Laboratoire associé au Centre National de la Recherche Scientifique.

# 1 Introduction

The baryon number to entropy ratio in the observed part of the universe is estimated to be $n_B/s \sim (4-6)10^{-11}$[1] . In 1967, A.D. Sakharov [2] established the three building blocks required from any candidate theory of baryogenesis: a) Baryon number violation, b) C and CP violation, c) Departure from thermal equilibrium.

The Standard Model (SM) contains a)[3] and b)[4], while c) could also be large enough [5][6], if a first order $SU(2) \times U(1)$ phase transition took place in the evolution of the universe [7]. An explanation within the SM would be a very economical solution to the baryon asymmetry puzzle. Unfortunately, intuitive arguments lead to an asymmetry many orders of magnitude below observation [10][11]. However, the study of quantum effects in the presence of a first order phase transition is rather delicate, and traditional intuition may fail. The authors of ref.[12] have recently studied this issue in more detail and claim that, at finite temperature, the SM is close to produce enough CP violation as to explain the observed $n_B/s$ ratio. In this letter, we summarize our quantitative study [8][9] of the Standard Model C and CP effects in an electroweak baryogenesis scenario. Even if one assumes an optimal sphaleron rate and a strong enough first order phase transition, we discard this scenario as an explanation of the observed baryon number to entropy ratio.

A first order phase transition can be described in terms of bubbles of "true" vaccuum (with an inner vaccuum expectation value of the Higgs field $v \neq 0$) appearing and expanding in the preexisting "false" vaccuum (with $v = 0$ throughout). We can "zoom" into the vicinity of one of the bubbles. There the curvature of its wall can be neglected and the world is divided in two zones: on the left hand side, say, $v = 0$; on the right $v \neq 0$ and masses appear. The actual bubble expands from the broken phase ($v \neq 0$) towards the unbroken one ($v = 0$). We work in the wall rest frame in which the plasma flows in the opposite direction. Consider thus a baryonic flux hitting the wall from the unbroken phase. Far enough to the left no significant CP-violating effect is possible as all fermions are massless. In consequence, the heart of the problem lies in the reflection and transmission properties of quarks bumpimg on the bubble wall. CP violation distinguishes particles from antiparticles and it is *a priori* possible to obtain a CP asymmetry on the reflected baryonic current, $\Delta_{CP}$. The induced baryon asymmetry is at most $n_B/s \sim 10^{-2} \Delta_{CP}$, in a very optimistic estimation of the non-CP ingredients [13][12].

The symmetries of the problem have been analyzed in detail [8] for a generic bubble. The analytical results correspond to the thin wall scenario. The latter provides an adequate physical description for typical momentum of the incoming particles $|\vec{p}|$ smaller than the inverse wall thickness $l$, i.e., $|\vec{p}| \ll 1/l$. For higher momenta, cutoff effects would show up, but it is reasonable to believe that the thin wall approximation produces an upper bound for the CP asymmetry.

The precise questions to answer in the above framework are : 1) the nature of the physical process in terms of particles or quasi-particles responsible for CP violation, 2) the order in the electroweak coupling constant, $\alpha_W$, at which an effect first appears, 3) the dependence on the quark masses and the nature of the GIM cancellations involved.

We shall consider the problem in two steps: zero temperature scenario in the presence of a wall ($T = 0$) and finite temperature case. Intuition indicates that an existing CP violating effect already present at zero temperature will diminish when the system is heated because the effective v.e.v. of the Higgs field decreases and in consequence the fermion masses do as well (only the Yukawa couplings already present at $T = 0$ remain unchanged). This intuition can be misleading only if a new physical effect, absent at $T = 0$ and relevant for the problem, appears at finite temperature. We discuss and compare the building blocks of the analysis in both cases. The $T = 0$ case provides a clean analysis of the novel aspects of the physics in a world with a two-phase vacuum. At $T = 0$ the quantum mechanical problem is well defined and the definition of particles, fields, in and out states, etc. is transparent.

At finite temperature, a plasma is an incoherent mixture of states. CP violation, being a typical quantum phenomenon, can only be observed if a high degree of strong <u>and</u> electroweak coherence is present. This is however not the case in the plasma, where the scattering of quasiparticles with thermal gluons induces a large damping rate.

The results of our analysis indicate that in the presence of a first order phase transition, a CP-violating baryon asymmetry in the SM appears at order $\alpha_W^2$, has a conventional type of GIM cancellation and chiral limit, and it is well below what observation requires in order to solve the baryon asymmetry puzzle.

## 2    Zero temperature

The necessary CP-odd couplings of the Cabibbo-Kobayashi-Maskawa (CKM) matrix are at work. Kinematic CP-even phases are also present, equal for particles and antiparticles, which interfere with the pure CP-odd couplings to make them observable. These are the reflection coefficients of a given particle hitting the wall from the unbroken phase. They are complex when the particle energy is smaller than its (broken phase) mass. Finally, as shown below, the one loop self-energy of a particle in the presence of the wall cannot be completely renormalized away and results in physical transitions. Such an effect is absent for on-shell particles in a world with just one phase. The difference is easy to understand: the wall acts as an external source of momentum in the one-loop process. The transitions between any two flavors of the same charge produce a CP violating baryonic flow for any given initial chirality.

The essential non-perturbative effect is the wall itself. The propagation of any particle of the SM spectrum should be exactly solved in its presence. And this we do for a free fermion, leading to a new Feynman propagator which replaces and generalizes the usual one. With this exact, non-perturbative tool, perturbation theory is then appropiate in the gauge and Yukawa couplings of fermions to bosons, and the one loop computations can be performed. Strictly speaking the gauge boson and Higgs propagators in the presence of the wall are needed, and it is possible to compute them with a similar procedure [14]. In particular this implies to consider loops with unbroken, broken and mixed contributions. For the time being, we work in a simplified case in which the wall does not act inside quantum loops. These are computed in the broken phase. We start from the following unperturbed Hamiltonian for a Dirac particle in the wall rest frame:

$$H = \vec{\alpha} \cdot \vec{p} + \beta m \theta(z) \qquad (1)$$

The propagator for quarks in the presence of the wall contains massless and massive poles:

$$S(q^f, q^i) = -1/2 \Big\{ \frac{1}{q_z^f - q_z^i + i\epsilon} \left( \frac{1}{\slashed{q}^f} + \frac{1}{\slashed{q}^i} \right) - \frac{1}{q_z^f - q_z^i - i\epsilon} \left( \frac{1}{\slashed{q}^f - m} + \frac{1}{\slashed{q}^i - m} \right) +$$

$$\frac{1}{\slashed{q}^f - m} \gamma_z \frac{1}{\slashed{q}^i} - \frac{1}{\slashed{q}^f} \gamma_z \frac{1}{\slashed{q}^i - m} -$$

$$\frac{m}{\slashed{q}^f(\slashed{q}^f - m)} \left[ 1 - \frac{m\gamma_0}{E + p'_z}(1 - \alpha_z) \right] \gamma_0 \frac{m}{\slashed{q}^i(\slashed{q}^i - m)} \Big\} \qquad (2)$$

where we have assumed for simplicity zero momentum parallel to the wall ($q_x^i = q_y^i = q_x^f = q_y^f = 0$). Due to the wall the initial and final $z$ components of the momentum need not be equal. All denominators in the usual Feynman propagators in eq.(2) should be understood as containing a supplementary $+i\epsilon$ factor. Besides this traditional source of phases, the propagator contains new CP-even ones in $p'_z = \sqrt{E^2 - m^2}$, which becomes imaginary in the case of total reflection ($E < m$ where $E$ is the fermion energy). The tree level reflection matrix for a massless Dirac fermion hitting the wall from the unbroken phase is given by

$$R = \frac{m\gamma_z}{p_z + p'_z}, \qquad (3)$$

the Dirac structure of which reflects the opposite chirality of the incoming and reflected state.

The reflection amplitude should vanish in the chiral limit as some positive powers of $m_i$ and $m_f$. When the internal loop is computed just in the unbroken phase, it should behave as

$$I_\mu(q_\mu), \propto \slashed{q} \qquad (4)$$

due to Lorentz covariance. The insertion of eq (4) on the quark propagation gives a null result whenever reflection occurs just once, either before or after the loop insertion, due to the action of the Dirac operator on a massless fermion. A non-trivial dependence of the amplitude on both $m_i$ and $m_f$ is needed, and the total effect should go to zero as some positive power of both masses, with an odd overall dependence on fermion masses since chirality flips upon reflection. The argument can be generalized to the situation where the broken phase is considered inside the loop, the difference with eq. (4) being terms independent of $\slashed{\phi}$ but proportional to the external masses, as will be shown below.

Next, it is possible to argue the type of GIM cancellations for quark masses inside the loop. Consider the amplitude for reflecting an initial quark $i$ into a final one $f$. The relevant terms correspond to the interference of diagrams with two different internal quark masses, $M$ and $M'$, to be summed upon. Each individual diagram is proportional to[3]

$$A(i \to f) \propto a + \frac{M^2}{M_W^2}(b + c \log \frac{M}{M_W}) + \frac{M^4}{M_W^4}(d + e \log \frac{M}{M_W}) + ..... \qquad (5)$$

Due to KM unitarity, the CP observable has to be proportional to an antisymmetric function of $M$ and $M'$ and must show a different functional dependence on $M$ and $M'$: typically $\sim \frac{M^2 M'^2}{M_W^4} \log(M^2)$. In practice, as shown below, it will behave as $\frac{M^2 M'^4}{M_W^6} \log M'^2$, i.e. a further mass suppression. It is important to notice that whenever only the unbroken phase is considered inside the electroweak loop, $c = d = e = 0$, because the fermionic mass dependence stems from pure Yukawa couplings. No antisymmetric function in $M, M'$ is viable, and the effect should vanish at order $\alpha_W^2$ in total rate. There is no reason, though, to neglect the broken phase, and we expect an $O(\alpha_W^2)$ contribution. The above considerations apply as well for a wall of thickness $l \neq 0$.

For a thin wall the non-local character of the internal loop is important because large particle momenta $\sim M_W$ are present and $l \ll M_W^{-1}$. Our calculation suggests that an even smaller result (although always at the same electroweak order) would follow for a more realistic thick wall, $l \gg M_W^{-1}$, where a local appoximation could be pertinent.

We renormalize by substraction at $q^2 = 0$. In the region $m_i, m_f, E \ll M_W$, the flavor changing contribution [8] to the CP asymmetry is[4]:

$$\Delta_{CP}^{T=0} = \sum_{i,f} \sum_{M,M'} [|A(i_R \to f_L)|^2 - |A(\bar{i}_L \to \bar{f}_R)|^2] = \left[\alpha_W \frac{\pi}{16}\right]^2 (-c_1 c_2 c_3 s_1^2 s_2 s_3 s_\delta)$$

---

[3]This dependence is in general a complicated function. For transparency, only the behaviour for $M < M_W$ is described here. As the internal loop is convergent in $M$, no pure $\log M$ terms can appear.

[4]We did not compute the flavor diagonal $O(\alpha_W^2)$ contribution. They would be of the same order.

$$\sum_{i\neq f}\sum_{l'=l+1}(-2)\sum_{k=j+1}S_{jk}(M_l,M_{l'})b_{jk}(E,m_i,m_f), \quad (6)$$

with $k = j+1$ modulo 4 and $l' = l+1$ modulo 3, and where the function $S$ contains the dependence on the quark masses inside the loop

$$S_{jk}(M,M') = I_j(M)I_k(M') - I_j(M')I_k(M). \quad (7)$$

The integrals $I_{1,2,3,4}$ are defined as

$$I_n(M) = \delta_n \int_0^1 dx \left[\frac{x(1-x)}{x + \frac{M^2}{M_W^2}(1-x)}\right]^{\gamma_n} x^{\beta_n} \quad (8)$$

with $\delta_1 = 3\delta_3 = M^2/M_W^2$, $\delta_2 = 3\delta_4/2 = 2 + M^2/M_W^2$, $\beta_n = (0,1,0,1)$ and $\gamma_n = (1,1,3/2,3/2)$.

The functions $b_{jk}$ are antisymmetric in $j,k$. They contain the mass dependence for the incoming and outgoing quarks,

$$b_{12}(E,m_i,m_f) = b_{23}(E,m_i,m_f) = -b_{13}(E,m_i,m_f) = 0.$$

$$b_{13}(E,m_i,m_f) = -\left(\frac{m_i m_f}{M_W^2}\right)^2 \frac{m_f^2}{M_W} \frac{|2E + p_z'^f - p_z'^i|^2}{|E + p_z'^f|^2} Re\left(\frac{1}{p_z'^i + p_z'^f}\right)$$

$$b_{14}(E,m_i,m_f) = +\left(\frac{m_i m_f}{M_W^2}\right)^2 \frac{m_f^2}{M_W} \frac{1}{|E + p_z'^f|^2} Re\left(\frac{(2E + p_z'^f - p_z'^i)(E + p_z'^{f*} - p_z'^{i*})}{p_z'^i + p_z'^f}\right)$$

$$b_{24}(E,m_i,m_f) = -b_{14}(E,m_i,m_f)$$

$$b_{34}(E,m_i,m_f) = \left(\frac{m_i m_f}{M_W^2}\right)^2 \frac{m_f^2}{M_W^2} \frac{E}{|E + p_z'^f|^2} Im\left(p_z'^f - p_z'^i\right). \quad (9)$$

Notice that every $b_{jk}$ vanish when either $m_i$ or $m_f$ go to zero, as we argued they should. All $b_{jk}$ but $b_{12}$ depend on CP-even phases associated with the non-locality of the internal loop, as well as on the tree level CP-even phases in eq.(3). $b_{12}$ depends only on the latter and could be $\neq 0$ only when $m_i, m_f$ and $E$ are not neglected in front of $M_W$. In each sector of quark charges the heavier masses will dominate the effect, as expected. Notice that a two-threshold structure is present, corresponding to $m_i, m_f$.

In the numerical results we use the exact values of the functions $S_{jk}(M,M')$. It is instructive, though, to show a fit for the particular case $M, M' \ll M_W$, appropiate for all quarks but the top:

$$S_{1,2}(M,M') \to \frac{M^2 M'^2}{M_W^4}\left[\frac{M^2}{M_W^2}\log\frac{M^2}{M_W^2} - \frac{M'^2}{M_W^2}\log\frac{M'^2}{M_W^2}\right] \quad (10)$$

and the ratio of the remaining $S_{j,k}(M,M')$ to the result in eq. (10) is given by $S_{1,3} \to -\frac{7}{15}$, $S_{1,4} \to +1.5$, $S_{2,3} \to -\frac{1}{30}$, $S_{2,4} \to -\frac{1}{4}$, and $S_{3,4} \to +\frac{1}{6}$, in the same limit. The total asymmetry, for "down" type quarks, is dominated by the $s_R \to b_L$ contribution, with a maximum value $5\,10^{-26}$ around the $m_b$ threshold. For "up" type quarks, we get a dominant $c_R \to t_L$ contribution of order $\sim 2.5\,10^{-25}$. $\Delta_{CP}^{T=0}$ is therefore many orders of magnitude smaller than needed for baryon number generation. Notice that the argument following eq.(4) fails when the wall itself is considered inside the quantum loops, as Lorentz covariance is lost, and the external factor $(\frac{m_i m_f}{m_W^2})^2$ could disappear, enhancing the effect for certain flavors, although leaving unchanged the qualitative conclusion. We consider the result of the present academic model as indicative of the electroweak order at which the effect first shows up, and of the well-behaved type of GIM cancellations of the physical process under study, which shows a standard chiral limit. It also helps to elaborate some non-pertubative tools for the physical $T \neq 0$ case.

## 3 Non zero temperature

The three building blocks are analogous to the $T = 0$ case: CKM CP violation, CP-even phases in the reflection coefficients[12] and the fact that the fermion self-energy at finite $T$ results in physical transitions. Let us consider the differences and new elements.

The argument developed after eq. (4) fails at finite temperature because Lorentz invariance is lost. A second difference stems from renormalization. At $T = 0$ we renormalize by subtraction at $q^2 = 0$. The remaining contribution is thus proportional to the virtuality, which is of the order of the fermion masses. At finite $T$, the additional thermal effects are not renormalized away, and the virtuality is of the order of the thermal masses $\sim g_s T \sim 50$GeV. Some powers of the mass that appeared at $T = 0$ may be replaced by powers of $T$.

At $T = 0$, far to the left in the unbroken, massless world, where the existence of the wall can be neglected, there is freedom to perform any Cabibbo-like rotation and thus to choose a basis. Such a freedom is lost at finite temperature [12] because the spectrum is not degenerate in both phases even far from the wall.

A fundamental difference with the $T = 0$ case is the damping rate, $\gamma$, of quasi-particles in a plasma. Due to incoherent thermal scattering with the medium, their energy and momentum are not sharply defined, but spread like a resonance of width $2\gamma$ [15]. The quasi-particle has thus a finite life-time, turning eventually into a new state, out of phase with the initial one. The QCD damping rate at zero momentum is of the order $\gamma \sim 0.15 g_s^2 T$ [15], i.e. $\sim 19$ GeV at $T = 100$ GeV. Intuitively, a large damping rate would kill the CP asymmetry, not unlike what happens, for example, in $D\bar{D}$ mixing. Although the imaginary part of the QCD self-energy is smaller than its real part, which settles the overall scale of the quasi-particle "masses", it is much larger than the real part of the electroweak self-energy. It should weaken the effect of electroweak level splitting, essential to the asymmetry.

A first step is the computation of the spectrum far from the wall. The dispersion relations describing it are given by the zeros of the self-energy,

$$\Sigma(\omega, k) = \omega + \vec{\alpha} \cdot \vec{k} + m_i \gamma_0 - Re(\delta\Sigma) - i\, 2\gamma \tag{11}$$

where $2i\gamma$ is the imaginary part, approximated here by its pure QCD component. The QCD contribution to $Re(\delta\Sigma)$ is described in refs.[16]-[19]. We compute the full SU(3)×SU(2)×U(1) real part of the thermal self-energy at the one loop level[5].

$$Re(\delta\Sigma) = -h(\omega, k) - a(\omega, k)\vec{\alpha} \cdot \vec{k} + c(\omega, k) m_i \gamma_0. \tag{12}$$

In the plasma rest frame and the mass basis,

$$a(\omega, k) = f\, A(m_i, 0) + \frac{g^2}{2}[\sum_l f_{W,l}\, A(M_l, M_W) + f_Z\, A(m_i, M_Z) + f_H\, A(m_I, M_H)] \tag{13}$$

with

$$f = [\tfrac{4}{3}g_s^2 + Q_i^2 g^2 s_W^2](L+R) \quad , \quad f_{W,l} = [(1+\tfrac{\lambda_i^2}{2})L + \tfrac{\lambda_i \lambda_f}{2} R] K_{li} K_{fl}^*$$
$$f_Z = \tfrac{1}{2}(\tfrac{4}{c_W^2}(T_i^3 - Q_i s_W^2)^2 + \tfrac{\lambda_i^2}{2})L + (\tfrac{4}{c_W^2}(-Q_i s_W^2)^2 + \tfrac{\lambda_i^2}{2})R \quad , \quad f_H = \tfrac{\lambda_i^2}{4}(L+R)$$
$$A(M_F, M_B) = \tfrac{1}{k^2} \int_0^\infty \tfrac{dp}{8\pi^2}([-\tfrac{(\omega^2+k^2+\Delta)}{2k}\tfrac{p}{E_B}L_I^+(p) - \tfrac{\omega p}{k}L_I^-(p) + \tfrac{4p^2}{E_B}]n_B(E_B) +$$
$$[\tfrac{\omega^2-k^2-\Delta}{2k}\tfrac{p}{E_F}L_{II}^+(p) - \tfrac{\omega p}{k}L_{II}^-(p) + \tfrac{4p^2}{E_F}]n_F(E_F)) \tag{14}$$

where
$$L_I^\pm(p) = [\log\frac{2kp + 2E_B\omega + \omega^2 - k^2 + \Delta}{-2kp + 2E_B\omega + \omega^2 - k^2 + \Delta}] \pm [E_B \to -E_B] \tag{15}$$

and $L_{II}^\pm(p)$ is obtained from eq. (15) with the replacements $E_B \to -E_F$, $\Delta \to -\Delta$ and a global minus sign. The remaining variables are $E_{F,B} = \sqrt{p^2 + M_{F,B}^2}$, $\Delta = M_B^2 - M_F^2$ and $n_{F,B}(E) = (\exp E/T \pm 1)^{-1}$.

The function $h(\omega, k)$ differs from the expression for $a(\omega, k)$, eq. (13), by a global minus sign and the replacement of the integral $A(M_f, M_B)$, eq. (14), by

$$H(M_F, M_B) = \frac{1}{k}\int_0^\infty \frac{dp}{8\pi^2}([pL_I^-(p) + \frac{\omega p}{E_B}L_I^+(p)]n_B(E_B) + pL_{II}^-(p)n_F(E_F)). \tag{16}$$

The chirality breaking term $c(\omega, k)$ is given by

$$c(\omega, k) = 2f\, C(m_i, 0) + \frac{g^2}{2}\left[\sum_l \frac{M_l}{m_i} g_{W,l}\, C(M_l, M_W) + g_Z\, C(m_i, M_Z) - f_H\, C(m_I, M_H)\right] \tag{17}$$

---

[5]$h(\omega, k) = \omega a(\omega, k) + b(\omega, k)$ in the notation of ref. [18]

with

$$g_{W,l} = \frac{\lambda_M}{2}(\lambda_f L + \lambda_i R) K_{li} K_{fl} \quad , \quad g_Z = \frac{1}{2}[-8Q_i s_W^2(T_3 - Q_i s_W^2) + \frac{\lambda_i^2}{2}](L+R)$$
$$C(M_F, M_B) = \frac{1}{k} \int_0^\infty \frac{dp}{8\pi^2} (\frac{p}{E_B} L_I^+(p) n_B(E_B) + \frac{p}{E_F} L_{II}^+(p) n_F(E_F)) \quad (18)$$

In the unbroken world the above expressions apply with all masses equal to zero.

The spectrum is well approximated in the "linear" region[6], i.e. for momentum $k$ such that $k \ll \omega$, by the effective Lagrangian:

$$\mathcal{L}_{eff} = \Psi_L^\dagger(i\partial_t - iA_L\vec{\partial}\cdot\vec{\sigma} - \omega_L)\Psi_L + \Psi_R^\dagger(i\partial_t + A_R i\vec{\partial}\cdot\vec{\sigma} - \omega_R)\Psi_R$$
$$+ i\Psi_L^\dagger \Gamma_L \Psi_L + i\Psi_R^\dagger \Gamma_R \Psi_R - (\Psi_L^\dagger \mu \Psi_R + \Psi_R^\dagger \mu^\dagger \Psi_L)\theta(z) \quad (19)$$

where $\Psi_L$ and $\Psi_R$ are respectively the left-handed and right-handed fields for either up or down type quarks. They are vectors in the three-dimensional flavour space. $\omega_{L,R}$, $A_{L,R}$, $\mu$ and $\Gamma_{L,R}$ are non-diagonal matrices in flavor space. We will neglect their $z$ dependence. $\mathcal{L}_{eff}$ is not hermitean due to the damping rate.

We solve the spectrum in pure $QCD$ and then perform an expansion in $\alpha_W$. As the $QCD$ damping rate is much larger than the electroweak effects, this is a sensible approximation[7]. At zero order in $\alpha_W$, $\Sigma(\omega, k)_{fi}$ is flavor and left-right diagonal and $\omega_{L,R}$ at this order are the solutions of

$$\omega^{0\,ii}_{L,R} + h^{ii}_{L,R}(\omega^{0\,ii}_{L,R}, 0) = 0 \qquad i = 1 \text{ to } 3. \quad (20)$$

and[8]

$$A_{L,R} = \frac{1 + a_{L,R}(\omega^0_{L,R}, 0)}{s_{L,R}(\omega^0_{L,R}, 0)}, \qquad \mu = \frac{m(1 - c(\omega^0_R, 0))}{s_L^{1/2}(\omega^0_L, 0)\, s_R^{1/2}(\omega^0_R, 0)}, \qquad \Gamma_{L,R} = \frac{2\gamma}{s_{L,R}(\omega^0_{L,R}, 0)} \quad (21)$$

where $a_{L,R}$, $h_{L,R}$ correspond to the coefficients of the projectors $L, R$ in $a, h$, and

$$s_{L,R}(\omega^0_{L,R}, 0) = 1 + \frac{\partial h_{L,R}(\omega^0_{L,R}, 0)}{\partial \omega}. \quad (22)$$

A numerical estimate for down quarks gives a group velocity $A_L = A_R = 0.339$ and $s_{L,R}(\omega^0_{L,R}, 0) = 1.89$, close to the values $1/3$ and $2$, respectively, obtained with the unbroken loop when just the leading $T^2$ terms are considered [12][9]. As the numerical difference is of no practical consequence we will use the latter values in the following arguments, even at non-trivial order in $\alpha_W$.

---

[6]This should be a good approximation for light quarks, as the CP effects will diminish for higher momenta.

[7]For zero damping rate, the inclusion of the electroweak effects can shift the poles in a way that might enhance the asymmetry. Not for a large width. The "exact" computation, i.e. the numerical results without performing an expansion in $\alpha_W$, will be presented in [9].

[8]The notation L,R is kept here for later use in left-right asymmetric contributions.

[9]For $u$ and $c$ quarks we obtain $A_L = A_R = 0.346$ and $h_{L,R}(\omega^0_{L,R}, 0) = 1.88$, while for the top the results are 0.165 and 2.5, respectively.

In order to fix the notations, assume first $\gamma = 0$ and thus $\mathcal{L}_{eff}$ hermitean. Consider quarks with a positive spin projection along the $z$ axis, or equivalently a right incoming chirality. The Dirac equation resulting from eq. (19) is then solved for eigenstates of the energy which are superpositions of an incoming, a reflected and a transmitted plane wave. For an energy $\omega$ (real) the reflection matrix $r(\omega)$ satisfies [12]:

$$\frac{m^\dagger}{2} + r\frac{m}{2}r - \check{p}_L r + r\check{p}_R = 0 \qquad (23)$$

where $m$ is the mass matrix of the external quarks and

$$\check{p}_L = -(\omega - \omega_L^0 + \delta p_L), \qquad \check{p}_R = \omega - \omega_R^0 + \delta p_R. \qquad (24)$$

$\omega_L^0 = \omega_R^0$ contains only the gluon contribution and $\delta p_L$, $\delta p_R$ are of order $\alpha_W$. A CP asymmetry arises from the interference of the non-diagonal parts of $\delta p_L$ and $\delta p_R$. These have the general structure

$$\delta p_R = \alpha_W \lambda_i \lambda_f \sum_l K_{li} K_{fl}^* I_R(M_l^2), \qquad \delta p_L = \alpha_W \sum_l K_{li} K_{fl}^* I_L(M_l^2) \qquad (25)$$

where

$$I_R(M_l^2) = -\pi\, H(M_l, M_W)\frac{1}{s_{L,R}}, \qquad I_L(M_l^2) = \lambda_l^2 I_R(M_l^2). \qquad (26)$$

We expand the matrix $r$ in powers of $\alpha_W$: $r = r^0 + r^1 + r^2$, with

$$r_{ii}^0 = \frac{m_i}{(\check{p}_L^0 - \check{p}_R^0) - \sqrt{(\check{p}_L^0 - \check{p}_R^0)^2 - m_i^2}} \qquad (27)$$

and $\check{p}_L^0$ and $\check{p}_R^0$ are $\check{p}_L$ and $\check{p}_R$, eq. (24), computed to order 0 in $\alpha_W$.

Notice that the reflection coefficients (27) can become complex for certain energy ranges, corresponding to total reflection. These are the siblings of the ones we found at $T = 0$, eq.(3).

Consider now a non vanishing damping rate. The quasiparticles that will eventually reach the wall have been "created" somewhere in the unbroken phase, as a result of their last scattering on a thermal particle. The creation probability is computed by imposing that the average density of quasi-particles, $n_F(E)$, corresponds to the Boltzmann law. The reflection density near the wall is given by the ratio of the reflected to the incoming flux at $(z = 0, t = 0)$:

$$n_r(z = 0, t = 0) = \int dE_0 \int_{-\infty}^0 dz_0\, dt_0\, |\psi_r(0, 0; z_0, t_0, E_0)|^2 N(z_0, t_0, E_0), \qquad (28)$$

which is an incoherent sum of the rate $N(z_0, t_0, E_0) = 2\gamma n_F(E_0)$ of quasiparticle creation at $(z_0, t_0)$ with average energy $E_0$, times the probability $|\psi_r|^2$ to find the quasi-particles near (0,0) after they reflected. $|\psi_r|^2$ decays like $e^{-2\gamma|t_0|}$.

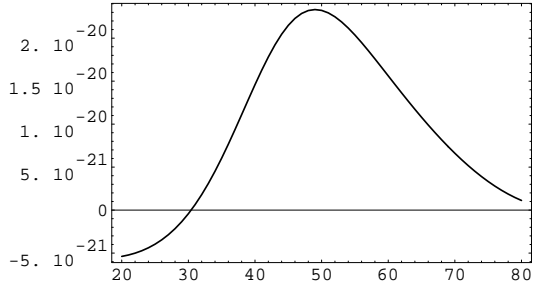

Figure 1: The dominant CP asymmetry when mass effects are included inside thermal loops, as a function of the energy. It corresponds to charge 2/3 flavors and appears at order $(O(\alpha_W^2))$.

It is adequate to describe the quasi-particles by wave packets with a small spatial extension $d$ relative to their mean free path $\sim 1/6\gamma$. Consider gaussian wave packets. One can show [9], using the analyticity of the functions $r_{ij}(\omega)$ and via a formal extension of the $t_0$-integral to $+\infty$, that

$$n_r(0,0) = \int dE_0\, n_F(E_0) \left[\int d\omega \frac{3d}{\sqrt{\pi}} e^{-9d^2(\omega-E_0)^2} |r(\omega+i\gamma)|^2 - \alpha m^2 (3d)^3 2\gamma\right] \tag{29}$$

for $m \ll \gamma \ll 1/3d$. $\alpha$ varies from 0 to $(\pi-2)/8\pi^{-1/2}$ depending on the importance of the would-be $t_0 > 0$ contribution[10]. The last term in eq. (29) can be neglected.

In this limit the reflected density is thus a gaussian smear-out of $|r(\omega+i\gamma)|^2$, with a maximum value $|r_{max}|^2 = m^2/16\gamma^2$, instead of 1 when $\gamma = 0$. One way to understand the physical origin of this reduction is to notice that, while the quasi-particles in the plasma are widely spread in energy and momentum, $d^{-1} \gg 6\gamma \gg m$, reflection (i.e. CP-even phases) is only important in a very narrow energy band, $\delta\omega \sim m$. Hence quasi-particles can hardly be reflected, but for the top flavor. In other words, it takes the wall a long time ($\sim 1/m$) to emit the reflected component of a small incoming packet. If the packet decays rapidly in a time $\sim 1/2\gamma$, it is natural to see the reflected wave strongly depleted by a factor $\sim m/2\gamma$. Furthermore, a CP asymmetry involves an interference between flavors and in consequence the relevant energy window is $\delta\omega \sim m_c, m_s$, respectively, for up and down sectors. The asymmetry for small wall velocity ($v_{wall}$) is $v_{wall}\Delta_{CP}$ [12], with

$$\Delta_{CP} = n_r(0,0) - \bar{n}_r(0,0) \simeq \int d\omega\, n_F(\omega) \Delta(\omega) \tag{30}$$

---

[10]This result is quite independent of the shape and width of the wall. The only requirement is the analyticity of $r(\omega)$ in the band $0 < \text{Im}(\omega) < 2\gamma$, verified for a thin wall, eq. (27).

and
$$\Delta(\omega) = \text{Tr} \left\{ r^\dagger(\omega + i\gamma) r(\omega + i\gamma) - \bar{r}^\dagger(\omega + i\gamma) \bar{r}(\omega + i\gamma) \right\} \quad (31)$$

where $\bar{r}$ and $\bar{n}$ refer to antiquarks.

The resulting asymmetry at order $\alpha_W^2$ is

$$\Delta(\omega) = \alpha_W^2 \left( -c_1 c_2 c_3 s_1^2 s_2 s_3 s_\delta \right) 4 b_{LR} S_{LR} \frac{|d_0|^2 + 2\text{Re}(d_0^2)}{|d_0|^4} \quad (32)$$

with $d_0 = -2i\gamma - 2\omega + \omega_R^0 + \omega_L^0$ and

$$\begin{aligned} S_{LR} &= \sum_l I_L(M_l^2) I_R(M_{l+1}^2) - I_R(M_l^2) I_L(M_{l+1}^2) \\ b_{LR} &= \frac{1}{i} \sum_i \lambda_i \lambda_{i+1} (r_{i+1\,i+1}^0 r_{ii}^{0*} - r_{ii}^0 r_{i+1\,i+1}^{0*}) \end{aligned} \quad (33)$$

with $l+1$ understood as modulo 3. $S_{LR}$ and $b_{LR}$ are the finite temperature analogs of the $T = 0$ results in eqs. (7) to (9). In eq.(32), higher powers of $m_f^2/(2\gamma)^2$ have been neglected (the numerical results below are computed without this approximation).

From the definitions (33) it is easy to check that the GIM mechanism is fully operative. In fact $S_{LR}/2$(respectively $b_{L,R}/2$) is the oriented area of a triangle built from the three internal (external) flavors, with coordinates $I_L + i I_R(M_l^2)$ ($\lambda_l r_{ll}^0$) in the complex plane, l=1,2,3. Notice that now the flavor-diagonal damping rate dominates over the flavor-dependent CP-even phases in the reflection matrix aligning these coordinates.

Using the following values for the masses in GeV, $M_W = 50$, $M_Z = 57$, $m_d = 0.006$, $m_s = 0.09$, $m_b = 3.1$, $m_u = 0.003$, $m_c = 1.0$, $m_t = 93.7$, the Yukawa couplings $\lambda_d = 1.2\,10^{-4}$, $\lambda_s = 1.8\,10^{-3}$, $\lambda_b = 6.2\,10^{-2}$, $\lambda_u = 6.2\,10^{-5}$, $\lambda_c = 2\,10^{-2}$ and $\lambda_t = 1.88$, and $\alpha_s = 0.1$, $\alpha_W = 0.035$ we obtain for the integrated asymmetry,

$$\frac{\Delta_{CP}^{uct}}{T} = 1.6\,10^{-21}, \qquad \frac{\Delta_{CP}^{dbs}}{T} = -3\,10^{-24}. \quad (34)$$

In both cases the asymmetry is dominated by the two heavier external quarks. The induced baryon asymmetry $n_b/s$ cannot exceed $10^{-2}$ times [12] these results.

Fig. 1 shows $\Delta(\omega)$ for up quarks.

In ref. [12] Farrar and Shaposhnikov (FS) obtain $\Delta_{CP}/T \gtrsim 10^{-8}$, and conclude $n_B/s \sim 10^{-11}$ (see eq. (10.3) in [12]). Their result is many orders of magnitude above ours, eq. (34). The main origin of the discrepancy is that they have not considered the effect of the damping rate on the quasi-particle spectrum[11]. A further difference is that they have considered only the unbroken phase inside the electroweak thermal loops. In this approximation a CP asymmetry appears first at order $\alpha_W^3$, while we get an effect at order $\alpha_W^2$. The reason has been discussed following eq.(5) for $T = 0$. Similarly at $T \neq 0$ one can see from eqs. (25) and (26), that when $M_l$=0, the right-handed contribution is flavor-diagonal (from $\sum_l K_{li} K_{fl}^* = \delta_{fi}$) forcing $\delta p_R$ to be diagonal and killing the $\alpha_W^2$ effect.

---

[11]More precisely, they take into account the finite mean free path of the quasi-particles in the suppression factor, i.e. what fraction of the $\Delta_{CP}$ is transformed into a baryon asymetry by the sphalerons, but not in the computation of $\Delta_{CP}$.

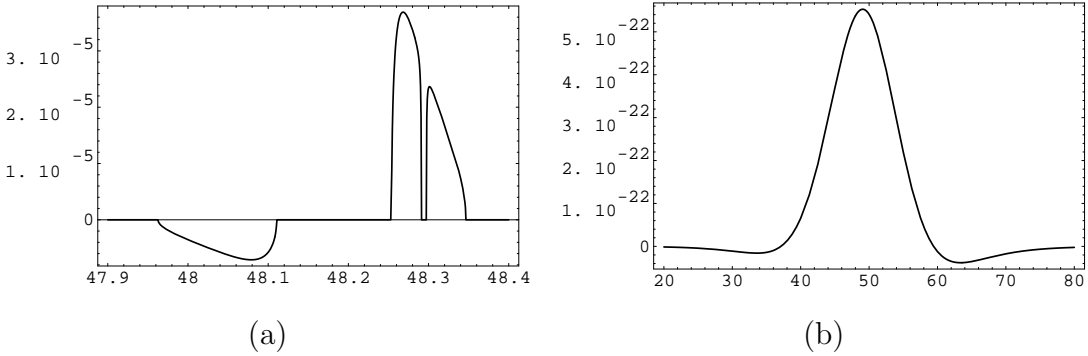

(a)                                     (b)

Figure 2: (a) shows the CP asymmetry produced by down quarks in the narrow energy range which dominates for zero damping rate, when masses are neglected in the internal loop. (b) shows the dramatic effect of turning on the damping rate effects, in the same approximation.

For the sake of comparison, we consider their approximation, i.e., with just the unbroken phase inside the thermal loops, both with zero and non zero damping rate, for a thin wall. In the energy region where the maximum asymmetry was found for $\gamma = 0$ [12] and down quarks, the $\alpha_W$ expansion with non zero damping rate leads to:

$$\Delta(\omega) = \left[\sqrt{\frac{3\pi}{2}} \frac{\alpha_W T}{32\sqrt{\alpha_s}}\right]^3 J \frac{(m_t^2 - m_c^2)(m_t^2 - m_u^2)(m_c^2 - m_u^2)}{M_W^6} \frac{(m_b^2 - m_s^2)(m_s^2 - m_d^2)(m_b^2 - m_d^2)}{(2\gamma)^9} \tag{35}$$

where $J = c_1 c_2 c_3 s_1^2 s_2 s_3 s_\delta$. This result shows the expected GIM cancellation and regular chiral behaviour. Its magnitude, $\sim 4\,10^{-22}$, is lower than the dominant one at order $\alpha_W^2$, shown in Fig. 2(b).

Furthermore, we confirm the validity of their numerical calculation with zero damping rate, with no $\alpha_W$ expansion involved and as can be seen in Fig. 2(a). The same computation including the damping rate is also shown in Fig.2(b).

A final comment on the wall thickness $l$ is pertinent. The mean free path for quasi-particles of lifetime $\sim \frac{1}{2\gamma}$ and group velocity $\sim \frac{1}{3}$ is $\sim \frac{1}{6\gamma} \sim \frac{1}{120} GeV^{-1}$. The thin wall approximation is valid only for $l \ll \frac{1}{6\gamma}$, while perturbative estimates[12] give $l \gtrsim \frac{1}{10} GeV^{-1} \gg \frac{1}{6\gamma}$. A realistic $CP$ asymmetry generated in such scenario will be orders of magnitude below the thin wall estimate in eq. (34), reinforcing thus our conclusions, because a quasi-particle would then collide and loose co-herence long before feeling a wall effect. This caveat should also be considered in any non-standard scenario of electroweak baryogenesis, where the wall thickness is larger than the mean free path.

We acknowledge Tanguy Altherr, Luis Alvarez-Gaumé, Alvaro De Rújula, Glennys Farrar, Jean Marie Frère, Jean Ginibre, Manolo Lozano, Jean-Yves Ollitreault, Anton Rebhan, Dominique Schiff and Misha Shaposhnikov for many inspiring discussions. We are specially indebted to Andy Cohen for stimulating discussions and we thank as well Carlos Quimbay for an excellent question. We admire the courage of Alvaro in his unsuccessful attempt to improve our writing.